\documentclass[pdflatex,sn-mathphys-num]{sn-jnl}

\usepackage{graphicx}%
\usepackage{multirow}%
\usepackage{amsmath,amssymb,amsfonts}%
\usepackage{amsthm}%
\usepackage{mathrsfs}%
\usepackage[title]{appendix}%
\usepackage{xcolor}%
\usepackage{textcomp}%
\usepackage{manyfoot}%
\usepackage{booktabs}%
\usepackage{algorithm}%
\usepackage{algorithmicx}%
\usepackage{algpseudocode}%
\usepackage{listings}%
\usepackage{svg}
\usepackage{float}
\usepackage{dirtytalk}
\theoremstyle{thmstyleone}%
\theoremstyle{thmstyletwo}%

\theoremstyle{thmstylethree}%

\begin{document}

\title[Article Title]{Diagnostic-free onboard battery health assessment}


\author[1]{\fnm{Yunhong} \sur{Che}}
\equalcont{These authors contributed equally to this work.}
\author[2,3]{\fnm{Vivek N.} \sur{Lam}}
\equalcont{These authors contributed equally to this work.}

\author[1]{\fnm{Jinwook} \sur{Rhyu}}

\author[1,4]{\fnm{Joachim} \sur{Schaeffer}}

\author[1]{\fnm{Minsu} \sur{Kim}}

\author[1,5]{\fnm{Martin Z.} \sur{Bazant}}

\author[2,3,6]{\fnm{William C.} \sur{Chueh}}

\author*[1]{\fnm{Richard D.} \sur{Braatz}}\email{braatz@mit.edu}

\affil[1]{\orgdiv{Department of Chemical Engineering}, \orgname{Massachusetts Institute of Technology}, \orgaddress{\postcode{02139}, \city{Cambridge}, \state{MA}, \country{USA}}}

\affil[2]{\orgdiv{Department of Materials Science and Engineering}, \orgname{Stanford University}, \orgaddress{\postcode{94305}, \city{Stanford}, \state{CA}, \country{USA}}}

\affil[3]{\orgdiv{Applied Energy Division}, \orgname{SLAC National Accelerator Laboratory}, \orgaddress{\postcode{94025}, \city{Menlo Park}, \state{CA}, \country{USA}}}

\affil[4]{\orgdiv{Control and Cyber-Physical Systems Laboratory}, \orgname{Technical University of Darmstadt}, \orgaddress{\postcode{64289}, \country{Germany}}}

\affil[5]{\orgdiv{Department of Mathematics}, \orgname{Massachusetts Institute of Technology}, \orgaddress{\postcode{02139}, \city{Cambridge}, \state{MA}, \country{USA}}}

\affil[6]{\orgdiv{Department of Energy Science and Engineering}, \orgname{Stanford University}, \orgaddress{\postcode{94305}, \city{Stanford}, \state{CA}, \country{USA}}}


\clearpage

\abstract{Diverse usage patterns induce complex and variable aging behaviors in lithium-ion batteries, complicating accurate health diagnosis and prognosis. Separate diagnostic cycles are often used to untangle the battery's current state of health from prior complex aging patterns. However, these same diagnostic cycles alter the battery's degradation trajectory, are time-intensive, and cannot be practically performed in onboard applications. In this work, we leverage portions of operational measurements in combination with an interpretable machine learning model to enable rapid, onboard battery health diagnostics and prognostics without offline diagnostic testing and the requirement of historical data. We integrate mechanistic constraints within an encoder-decoder architecture to extract electrode states in a physically interpretable latent space and enable improved reconstruction of the degradation path. The health diagnosis model framework can be flexibly applied across diverse application interests with slight fine-tuning. We demonstrate the versatility of this model framework by applying it to three battery-cycling datasets consisting of 422 cells under different operating conditions, highlighting the utility of an interpretable diagnostic-free, onboard battery diagnosis and prognosis model.


}




\maketitle
\section{Introduction}\label{sec1}

The growing trend of electrification has driven the wide application of lithium-ion batteries in electrified transportation and grid storage \cite{dunn2011electrical, schaeffer2024gaussian}. One of the primary concerns during battery usage is degradation, leading to reduced capacity and power capabilities. Knowing a battery's current and future state of health (SOH) is critical in optimizing its usage. However, the complex path-dependent nature of battery degradation results in diverse practical degradation patterns, posing challenges for accurate health diagnosis and prognosis \cite{ dubarry2018durability, schaeffer2024gaussian}. Typically, a diagnostic cycle is employed to assess the SOH of the battery \cite{edge2021lithium, LEWERENZ2019680, van2023interpretable, geslin2024dynamic, zhuang2024physics, rhyu2024optimum}. This diagnostic cycle may include pulse tests to check resistance, low-rate capacity checkups, and other application-specific tests necessary to extract useful SOH metrics. 

One common SOH metric is the low-rate battery capacity. However, this metric alone is insufficient as multiple degradation pathways can lead to the same cell-level SOH, limiting the interpretative power of solely relying on cell-level capacity. To address this, electrode-specific capacities have been proposed and extensively studied as a means to further breakdown degradation into electrode-level phenomena, capturing more intricate aging behavior of lithium-ion batteries \cite{van2023interpretable, geslin2024dynamic, mohtat2019towards, dey2019battery, CHE20231405}. This mechanistic model, known as differential voltage analysis (DVA), reconstructs the battery's open-circuit voltage (OCV) by subtracting matched and rescaled half-cell OCV curves. Using these half-cell curves, DVA reveals key modes of battery aging such as loss of active materials (LAM) and loss of lithium inventory (LLI) \cite{marongiu2016board, dubarry2012synthesize, stadlercombining}. While this analysis provides valuable in situ insights into aging mechanisms, it is typically performed on pseudo-OCV curves at rates of C/20 or lower taken during a diagnostic cycle. Furthermore, even at low rates, this adjustment does not fully capture the pseudo-OCV curve due to changes in the half-cell overpotential, inhomogeneities of lithiation, and other aging effects \cite{FATH2019100813, SCHMITT2022231296, SIEG2020101582}.

While diagnostic cycles provide valuable insights into the battery's SOH, such as enabling DVA, they have significant drawbacks. Taking time away from aging to perform disruptive and time-intensive diagnostic cycles can drastically alter the trajectory of battery aging \cite{LEWERENZ2019680}. From a practical point of view, time-consuming diagnostic tests cannot be performed routinely on consumer electronics, leading to the need for onboard diagnostics that use operational data. In real-world applications, battery charging current rates (C-rates), state-of-charge (SOC) ranges/voltage windows, and load profiles can vary significantly \cite{pozzato2023analysis, sulzer2021challenge, schaeffer2024gaussian}, necessitating data-driven models that are robust across diverse operating conditions. Moreover, a significant number of lithium-ion batteries are expected to retire from electric vehicles (EVs) in the near future, but performing a lengthy diagnostic cycle to evaluate their performance will be costly \cite{harper2019recycling, ciez2019examining, xu2020efficient}. To facilitate more intelligent repackaging and reuse, rapid evaluation of battery performance has garnered increased attention. Finally, different C-rates-based diagnostic tests are required in diverse applications, necessitating the model's suitability under varying scenarios.

In recent years, the advanced development of machine learning models has led to their implementation in battery health estimations and predictions \cite{schaeffer2024gaussian, roman2021machine, thelen2024probabilistic, severson2019data, nozarijouybari2024machine, hu2020battery}. Data-driven predictions of the OCV curve and incremental capacity (IC) curve have shown potential for onboard health diagnosis by using measured current and voltage \cite{tian2021electrode, tang2021reconstruction}. Deep neural networks are used predominantly in these data-driven models for the prediction of the battery OCV curve or charging curve due to their good nonlinear mapping capabilities \cite{HOFMANN2024100382, TIAN2021283, su2023battery, KO2024122488}. However, despite achieving high prediction accuracy, the input data is usually fixed to a full voltage curve under constant current between the voltage limits. This limits generalization to real applications with partial SOC cycling, or dynamic load profiles. Additionally, the ``black box" nature of these models limits their interpretability for a battery engineer to better understand the failure modes of a given battery. Consequently, mechanistic models are still required for subsequent aging mechanism analysis, separate from machine learning. 

To improve the interpretability of battery health models, promising ways for combining physical models with machine learning have been discussed \cite{aykol2021perspective, navidi2024physics}. One approach involves extracting physically interpretable features from mechanistic models to inform machine learning models used for battery modeling, health estimation, and lifetime prediction \cite{van2023interpretable, geslin2024dynamic, weng2021predicting, tu2023integrating}. However, extraction of these physical features poses challenges, and typically requires specific and time-consuming diagnostic tests. Physics-informed neural networks (PINNs), which integrate governing equations from mechanistic models to constrain machine learning models, show promise in improving both interpretability and the speed of solving governing equations in complex models \cite{karniadakis2021physics, borah2024synergizing, huang2024minn}. However, not only is constructing explicit models with partial derivatives challenging, but current models also do not fully capture highly interconnected degradation processes during battery aging \cite{wang2024physics}.

Here, we propose an interpretable model for the diagnosis and prognosis of complex battery aging that can be performed without separate diagnostic tests. Instead of a diagnostic, this model uses partial operating data with random SOC windows truncated from the full charging/discharging curves. An autoencoder integrates DVA-based mechanistic states in a physically meaningful latent space, enabling interpretable battery diagnosis and prognosis. Consequently, our model demonstrates the potential of machine learning to replace time-consuming and costly offline testing and model fitting, to enhance on-board health management without requiring historical data. Furthermore, by fine-tuning our model, we demonstrate the swift transferability to different applications. We evaluate the model for three datasets: the van Vlijmen et al.\  \cite{van2023interpretable} dataset, which comprises 236 batteries sourced from a Tesla Model 3 cycled under 126 different constant current operating conditions; the Geslin et al.\  \cite{geslin2024dynamic} dataset, which comprises 92 batteries under both constant and dynamic cycling conditions; and a new dataset consisting of a subset of 94 batteries from the van Vlijmen et al.\ dataset that underwent further low-rate (such as C/80 and C/40) validation tests. These exemplary use cases demonstrate the potential of diagnostic-free aging studies and quick onboard health assessment.

\section{Results}\label{sec2}
\subsection{Data generation}\label{subsec2-1}

\begin{figure}[H] 
    \centering 
    \includegraphics[width=\textwidth]{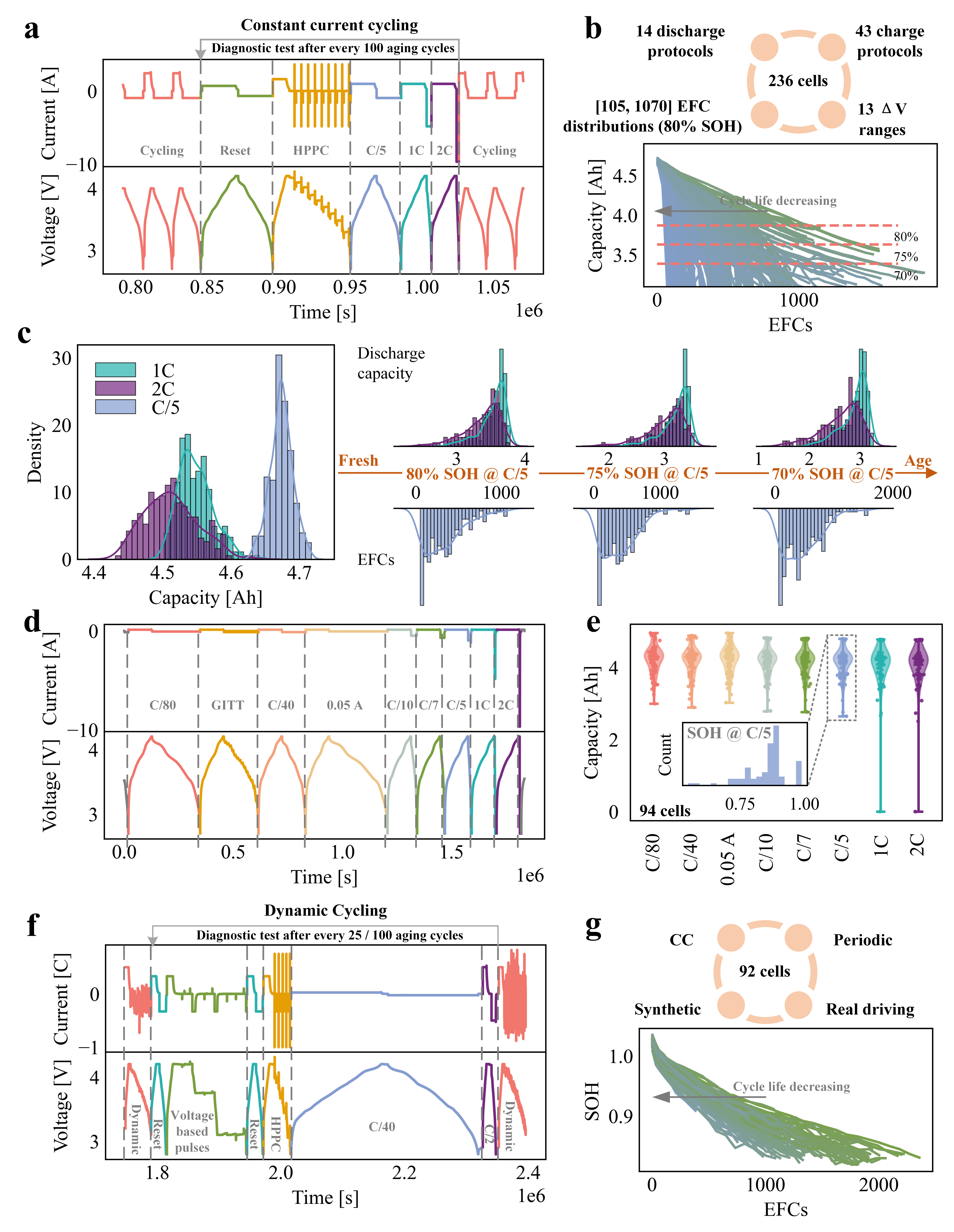}
    \vspace{-0.5cm}
    \caption{\textbf{Data illustration.} \textbf{a}) Protocol for battery aging in the van Vlijmen et al. dataset, which includes the cycling and diagnostic test. The cycling test uses different C-rates and SOC windows for the charging and discharging, while the diagnostic cycle includes a reset test, HPPC test, and RPT tests with three different C-rates. \textbf{b}) Summary of the testing conditions and the degradation curves of the test batteries. \textbf{c}) Capacity distributions obtained by different discharge C-rates and the variations during aging \cite{van2023interpretable}. \textbf{d}) Protocol for the different C-rates test generated in this work, where 8 different rate tests are included. \textbf{e}) Capacities distributions of the cells at different discharge C-rates. \textbf{f}) Protocol design for battery aging in the Geslin dataset \cite{geslin2024dynamic}, which includes the cycling and diagnostic test. \textbf{g}) Cycling design and the degradation curves for the 92 cells in the Geslin dataset \cite{geslin2024dynamic}.}
    \label{data_fig}
    \vspace{0.0cm} 
\end{figure}

Three datasets are evaluated in this work. Firstly, 236 cylindrical 21700 cells (Li(Ni,Co,Al)O$_\text{2}$/Graphite + SiO$_\text{x}$) extracted from a Tesla Model 3 EV were selected from the dataset presented in van Vlijmen et al.\  \cite{van2023interpretable}. These cells were cycled under 126 distinct operating conditions, including 43 charging protocols, 14 discharging protocols, and 13 voltage windows. This extensive range of test conditions produced complex and varied degradation curves with a broad distribution of equivalent full cycles (EFCs) at end of life. The typical procedure for battery aging is illustrated in Fig.~\ref{data_fig}a. The SOH was evaluated with a diagnostic test, which includes a reset cycle, hybrid pulse power characterization (HPPC), and reference performance test (RPT) at three C-rates (C/5, 1C, and 2C). The aging cycles consist of different charging and discharging protocols to cycle the cells within various voltage ranges for approximately 100 cycles before another diagnostic cycle is performed. Due to battery manufacturing variability and diverse aging conditions, various degradation patterns were observed, as depicted in Fig.~\ref{data_fig}b and details described in \cite{van2023interpretable}. The distribution of the 1C and 2C at fixed C/5 SOH  in Fig.~\ref{data_fig}c (See Note S1 for details) indicate that the mapping relationships between higher C-rates and reference RPT (C/5) are complex and diverse across different cycling conditions.

While the lowest rate used in the van Vlijmen dataset is C/5, lower C-rates testing such as C/40 are typically used for effective mechanistic feature extraction in aging analysis \cite{harlow2019wide, dubarry2009identify, fly2020rate, SCHMITT2022231296}. Obtaining these pseudo-OCV curves at different SOHs is time-consuming, yet essential to develop accurate state monitoring. To showcase the robustness of the model architecture when applied to a variety of C-rate cases for broader application interests, 94 cells from the van Vlijmen dataset \cite{van2023interpretable} at different SOH were subject to additional tests to generate a new dataset. An RPT was conducted using eight different C-rates as shown in Fig.~\ref{data_fig}d. The discharge capacities at different C-rates shown in Fig.~\ref{data_fig}e illustrate the large variations in RPT and mechanistic health status among these cells. In particular, these large variations highlight the difficulty in using high C-rate data to predict and understand low-rate data. Consequently, this dataset encompasses a wide range of aging stages, from fresh cells to retired ones, enabling a holistic study of battery aging over the entire lifespan and enabling validations for health diagnostics under different C-rates tests. For more details on the data samples, see Note S2. 

Finally, the van Vlijmen et al.\ dataset and the different C-rate validation dataset, use constant current (CC) cycling, however dynamic cycling is more representative of EV operations and leads to a prolonged lifetime compared to standard CC cycling \cite{geslin2024dynamic}. Therefore, model performance under operation closer to real-world aging is crucial. To this end, we use 92 cells from the Geslin et al.\ dataset \cite{geslin2024dynamic} as the third test case. The batteries were subject to four types of operating conditions i.e., CC, periodic, synthetic, and real driving protocols. A periodic diagnostic cycle is conducted after every 25/100 aging cycles. The typical profile and the degradation curves are shown in Figs.~\ref{data_fig}fg. During the diagnostic test, voltage-based pulses are included besides reset, HPPC, and RPT tests. The RPT was performed under C/40, providing data closer to real OCV for the verification of our model. The diversity of degradation during practical cycling further motivates the utility of onboard health assessment duringoperation.

\subsection{Health diagnosis}\label{subsec2-2}

\begin{figure}[H] 
    \centering 
    \includegraphics[width=\textwidth]{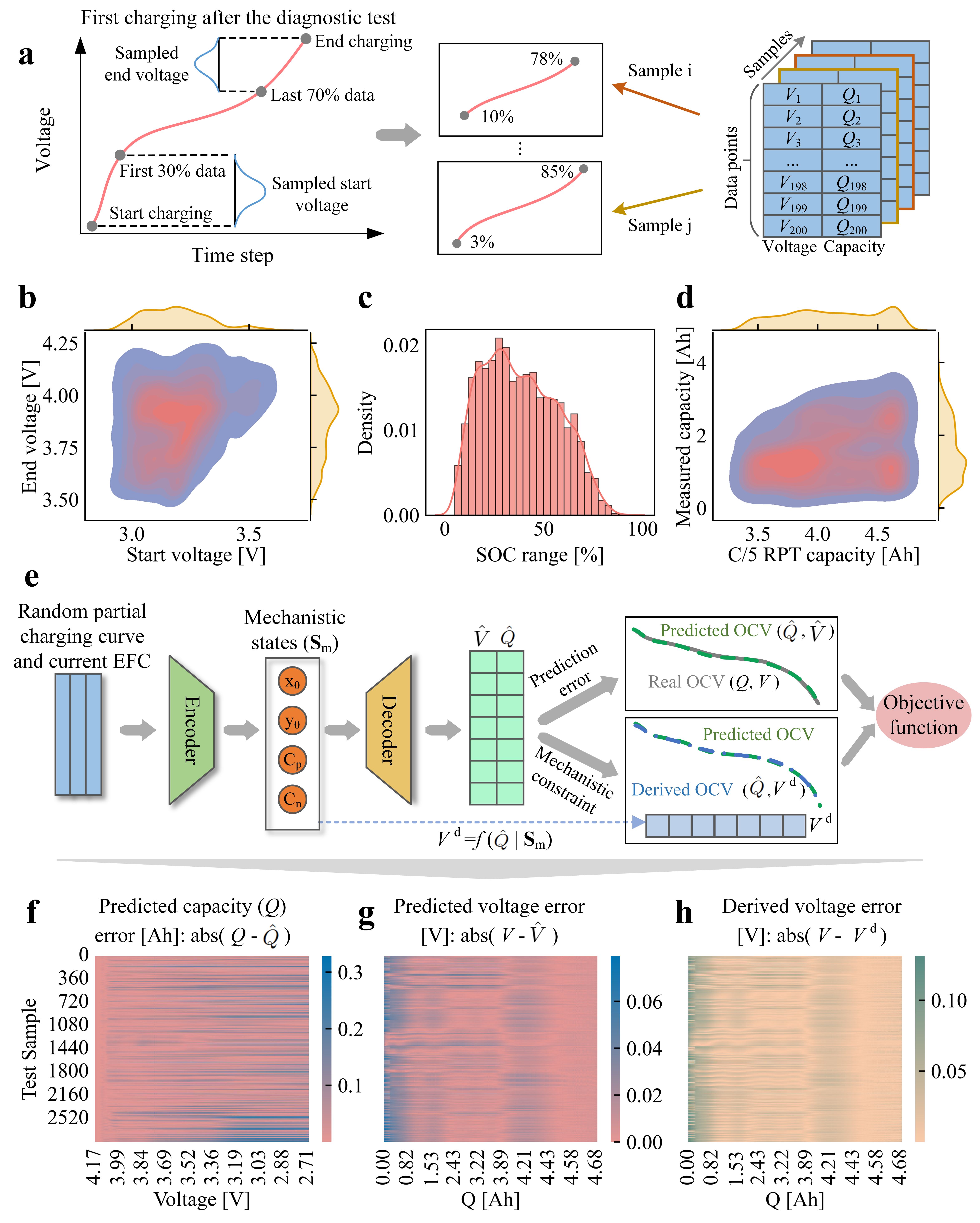}
    \vspace{-0.5cm}
    \caption{\textbf{Data sampling and diagnostic model.} \textbf{a}) Randomly sampled portions of charging curves. \textbf{b}) Distributions of the voltage windows for the input data sampling of the cycling dataset. \textbf{c}) The SOC ranges of the extracted partial charging data. \textbf{d}) The mapping between the measured capacity from the partial charging and the reference C/5 RPT capacity. \textbf{e}) Mechanistic constraint model for battery health diagnosis using randomly sampled portions of charging curves. \textbf{f--g}) The errors of the $\hat{Q}$ sequence and $\hat{V}$ sequence of the predicted OCV curves for all the test samples (composed of all the RPT cycles from the test cells), respectively. \textbf{h}) The errors of the derived voltage of the OCV curves, which are calculated based on the physics-constrained mechanistic states in the machine learning model. The predicted discharge capacity ($\hat{Q}$) and voltage ($\hat{V}$) MAE errors were 45.1 mAh and 10.1 mV, respectively, with derived voltages ($V^{\text{d}}$) showing MAE error below 18.4 mV.}
    \label{sample_fig}
    \vspace{0.0cm} 
\end{figure}

First, we evaluate our model on the van Vlijmen et al.\ dataset. The first charge (with rates between C/5 and 2 C) after the diagnostic cycle is used as the input to the model, and the output is the full C/5 discharge curve of the directly preceding diagnostic cycle, which will be used as the \textit{pseudo}-OCV curve in this study. In a real-world scenario, an EV battery may be charged between a variety of SOC ranges depending on their daily driving requirements. Although the SOC ranges and charging C-rates are already varied in the van Vlijmen et al.\ dataset to make the model robust to a large variety of working SOC ranges, the beginning and end of the charging data are artificially truncated as demonstrated in Fig.~\ref{sample_fig}a (see Note S1 for details), yielding several partial charging voltage curve samples that correspond to the same C/5 discharge curve for the purpose of model training. The truncated voltage and capacity, together with the current EFC, are employed as input measured data for the full OCV (C/5 discharge curve) predictions. Fig.~\ref{sample_fig}b demonstrates the distributions of the data samples with different voltage windows, and the corresponding SOC ranges are shown in Fig.~\ref{sample_fig}c. Based on prior studies that assessed daily usage conditions in real-world applications, where the obtainable SOC ranges could significantly vary depending on the applications and usage habits, our data sampling effectively simulates practical charging scenarios \cite{schaeffer2024gaussian, pozzato2023analysis, zhao2021assessment, figgener2024multi, cui2022battery, zhao2024challenges}. The relationship between partially charged capacities from randomly sampled portions of charging processes and the reference capacity (C/5 discharge capacity) is illustrated in Fig.~\ref{sample_fig}d, demonstrating a significant nonlinearity.

To go from the partial charging curves to the full C/5 discharge voltage curves, the encoder-decoder structure is employed for the data-driven health diagnosis as represented by Fig.~\ref{sample_fig}e. The encoder takes the partial charging voltage, capacity, and EFC and encodes them into 4 latent variables. These latent variables are assigned to be the mechanistic states used in DVA: anode capacity ($C_n$), cathode electrode capacity ($C_p$), anode ($x_0$) and cathode lithiation state ($y_0$) at the beginning of discharge of the corresponding half cell. These latent variables become these mechanistic states through the mechanistic constraint which compares the derived OCV from these latent variables to the predicted OCV (see Notes S3--S4 for detailed descriptions of the model and physics constraints). The decoder uses these mechanistic states and outputs the C/5 discharge curve. The objective function is a combination of the prediction error and the mechanistic constraint error. In summary, this model can be understood as being first, a flexible DVA where the encoder portion enables performing the analysis on partial charging data, followed by a decoder that enables flexible reconstructing of the voltage curve that balances pure DVA reconstruction (mechanistic constraint) and a component that adjusts the voltage curves to account for the discrepancy (prediction error).

To evaluate the model comprehensively, all cells with the same charging/discharging protocols appeared only in either the training or testing group. This setup ensured that the testing cells were aged under conditions unseen in the training data. Additionally, 20\% of the training data were randomly chosen as validation samples to prevent over-fitting by early stopping, thereby establishing an inner loop for validation and an outer loop for testing (see Table S3 for the detailed grouping information). Figs.~\ref{sample_fig}fg show the prediction results for the OCV curve, and the derived OCV (represents the OCV curve obtained directly by the constrained mechanistic states) is shown in Fig.~\ref{sample_fig}h. The heat maps reflect the capacity/voltage error at each voltage/capacity point for all the test OCV curves, where the color bars represent the absolute errors. The results indicate the stability of the physics-constrained machine learning model for health diagnosis across batteries operating under diverse C-rates, SOC ranges, and aging states. The mean absolute error (MAE) for the predicted discharge capacity ($\hat{Q}$) and voltage ($\hat{V}$) errors are 45.1 mAh (0.93 \% of the nominal capacity) and 10.1 mV, respectively, with derived voltages ($V^{\text{d}}$) with MAE of 18.4 mV, as summarized in Table~\ref{tab1}. Due to the variations in open circuit overpotential during aging \cite{SCHMITT2022231296}, we observe that while the predicted voltage error and the derived voltage error have a qualitatively similar error dependence on capacity, the predicted voltage errors (Fig.~\ref{sample_fig}g) are lower, indicating that machine learning can compensate for the deficiencies in the mechanistic model during aging.

Our model provides SOH estimation in each cycle and is comparable with different neural network architectures as demonstrated in Note S4. The prediction results indicate SOH estimation errors remained below 7\% under all conditions with more than 95\% of the results having relative errors below 5\% with a simple multilayer perception model as the encoder and decoder. The error distributions shown in Fig.~S9 indicate that the estimations are not sensitive to input voltage and SOC ranges. The performance underscores the high reliability of the model for SOH estimation across diverse application scenarios.

\begin{figure}[H] 
    \centering 
    \includegraphics[width=\textwidth]{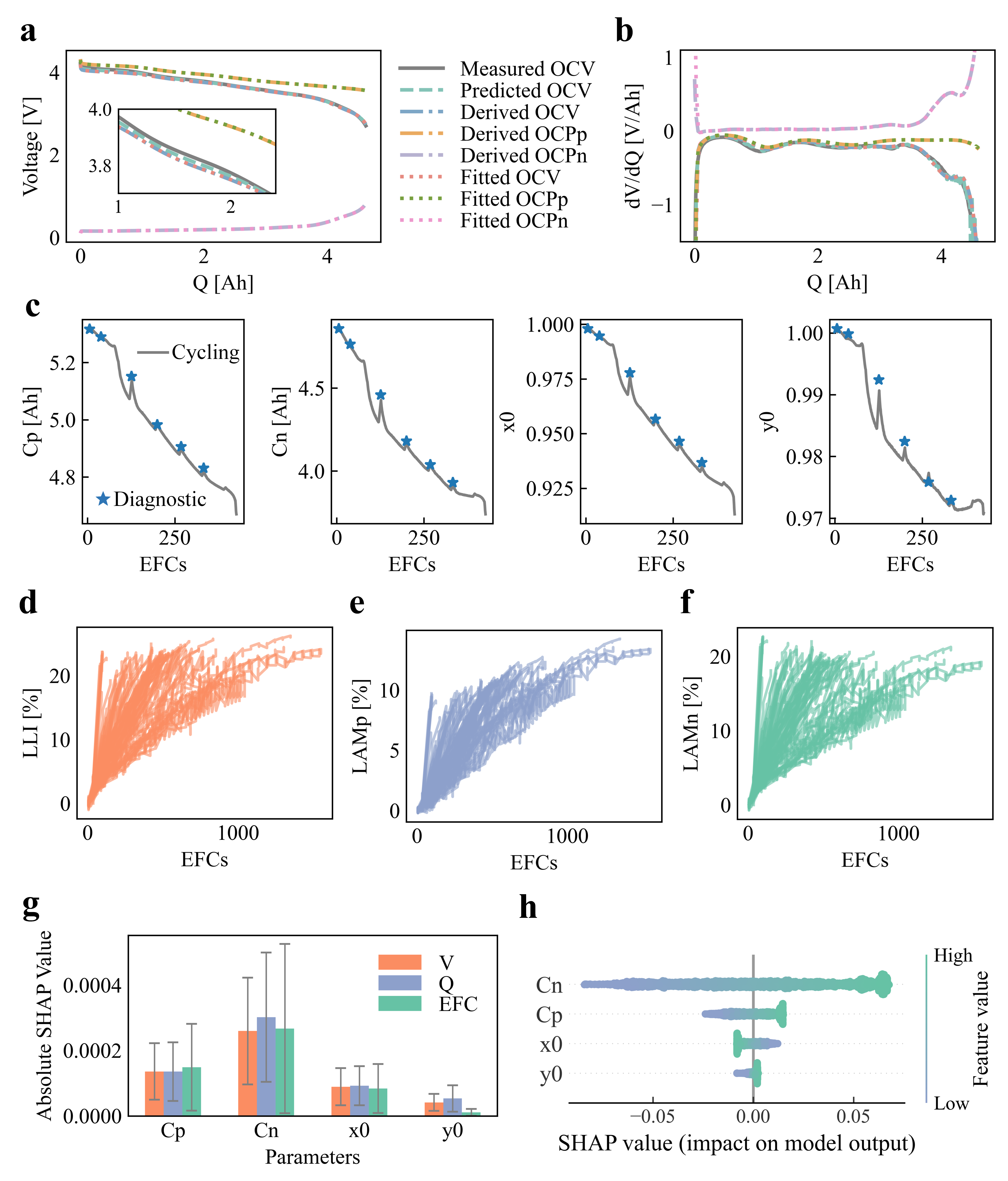}
    \vspace{-0.5cm}
    \caption{\textbf{Health diagnosis.} \textbf{a}) Illustration of the predicted OCV curves, derived OCV curves based on the constrained mechanistic states, and the fitted OCV based on the states obtained by the particle swamp optimization for the predicted OCV curves. \textbf{b}) The differential voltage curves for the corresponding curves in \textbf{a}. \textbf{c}) The predicted curves of the mechanistic states obtained from cycling cycles and the comparisons from those obtained from diagnostic cycles. \textbf{d--f}) The predicted aging modes, which are directly obtained based on the constrained mechanistic states, for onboard diagnosis of LLI, LAM at the cathode (LAM$_\text{p}$), and LAM at the anode (LAM$_\text{n}$). \textbf{g}) The SHAP analysis results for the encoder, which represents the impacts (absolute values) of the input information from the partial charging curve on the prediction of mechanistic states. \textbf{h}) The SHAP analysis of the decoder for the OCV curve predictions, where the impacts of the constrained mechanistic states on the predicted OCV (combination impacts of the capacity and voltage) are illustrated.}
    \label{diagnosis_interp}
    \vspace{0.0cm} 
\end{figure}

Exemplary prediction results for the C/5 pseudo-OCV and differential voltage curves are shown in Figs.~\ref{diagnosis_interp}ab and S10, demonstrating the effectiveness and physical interpretability of the model. The directly predicted OCV, derived OCV, and half-cell voltage facilitate the interpretation of holistic health status using only the information from the randomly sampled portions of the charging curve. The mechanistic states capture the aging and the decoder further compensates the prediction errors. To illustrate the role of mechanistic constraints in the data-driven model, we present example predictions without the physical constraint and the corresponding boundary constraint ($L_{\text{phy}}$ and $L_{\text{bound}}$ in Note S3) in Fig.~S11. Although a purely data-driven model can predict OCV curves based on partial charging data, it functions as a “black box” with non-physical meanings. Additionally, the derived OCV curve can extend beyond reasonable ranges without boundary constraint, preventing the interpretation of aging mechanisms in a purely data-driven model. Further results from the proposed model, including the worst and average diagnosis performances across all tests, are provided in Fig.~S12.

With this model, we can now generate the C/5 discharge voltage curves using the charging portion of operational data without a separate diagnostic cycle. 
In Fig.~\ref{diagnosis_interp}c,  we extract the mechanistic states from an example cell using solely the aging cycle data. For the complete prediction of the C/5 discharge curves for every aging cycle of this representative cell, see Fig.~S13. From this analysis, we can see how the full SOC range diagnostic cycle modifies the state of the battery and generates a capacity increase that is often observed in battery degradation studies \cite{LEWERENZ2019680, GUO202234}. This approach enables a full trajectory picture of degradation in the battery that is otherwise inaccessible. 

This approach is applied to all the testing cells in the dataset and from these mechanistic states we can directly obtain the degradation modes: loss of lithium inventory (LLI), and loss of active material on the cathode (LAM$_\text{p}$) and anode (LAM$_\text{n}$). Results in Figs.~\ref{diagnosis_interp}d--f illustrate the degradation of each aging mode derived from our model, with the variations of each mechanistic state shown in Fig.~S14. The results indicate that the anode degradation progresses more rapidly than the cathode degradation, leading to greater losses of active material on the anode. The lithium inventory also exhibits a rapid decline, as shown in the diagnosis results. The Li(Ni,Co,Al)O$_\text{2}$/Graphite + SiO$_\text{x}$ cells used for the aging test, have a strong coupling between LLI and LAM$_\text{n}$ (Fig.~S15). As the anode is a blended Graphite SiO$_\text{x}$ electrode, swelling of Si with Li intercalation can cause cracking of particles and disconnect, creating `dead' particles with trapped/lost Li \cite{Mikheenkova_2023, Liang_2023, KIRKALDY2024234185}. LAM$_\text{p}$ is also correlated with LLI as demonstrated in Fig.~S15. Possible causes are oxidation-induced cation disordering or cathode-electrolyte interphase (CEI) growth degrading the cathode while trapping dead Li in blocked crystal sites \cite{Zhuang_2022, Mikheenkova_2023, KIRKALDY2024234185}.

SHAP (SHapley Additive exPlanations) \cite{NIPS2017_8a20a862, mangalathu2020failure, cui2024data, geslin2024dynamic, van2023interpretable, lee2022state}, a tool for analyzing the importance of input features on the output, is used in combination with our model to further interpret our results. The SHAP analysis for the encoder is presented in Fig.~\ref{diagnosis_interp}g, while the results for the decoder are illustrated in Fig.~\ref{diagnosis_interp}h. For most of the mechanistic states, voltage, capacity, and EFC have approximately equal importance in determining the output. For the decoder, however, the anode capacity contributes most significantly to the reconstruction of the voltage curve. The aging mode analysis indicates a more pronounced degradation of LAM$_\text{n}$ compared to LAM$_\text{p}$, which is corroborated by the SHAP results. In addition, due to the fact the anode is a blended electrode the decoder may need to compensate for the voltage effects of preferential degradation of Si \cite{SCHMITT2022231296, BONKILE2024234256}. The analysis highlights that the capacity degradation on the anode has a greater impact on health predictions than the capacity degradation on the cathode, as depicted in Fig.~\ref{diagnosis_interp}h. Additional results detailing the impact of features on mechanistic state estimations can be found in Fig.~S16. We have shown that this model can enable diagnostic-free interpretable health assessment of batteries across various SOC ranges for charging applications.

\subsection{Health prognosis}\label{subsec2-3}

In addition to health diagnosis to understand current aging conditions, health prognosis is essential to anticipate future degradation to guide predictive maintenance and reduce research and development costs. Future degradation can be forecasted using the macroscopic (current EFC and predicted SOH at the current cycle, i.e., the last value of $\hat{Q}$ sequence) and mechanistic states ($C_\text{p}$, $C_\text{n}$, $x_0$, and $y_0$). The data-driven diagnostic results are utilized as the basis for future degradation prediction, without needing access to historical degradation data. This allows for developing a sequential framework for health diagnosis and prognosis (as detailed in Note S3). Such a framework facilitates real-world applications by eliminating the need for increased memory capacity in the battery management system for health prognosis, which conventional degradation prediction models typically require.  

\begin{figure}[H] 
    \centering 
    \includegraphics[width=\textwidth]{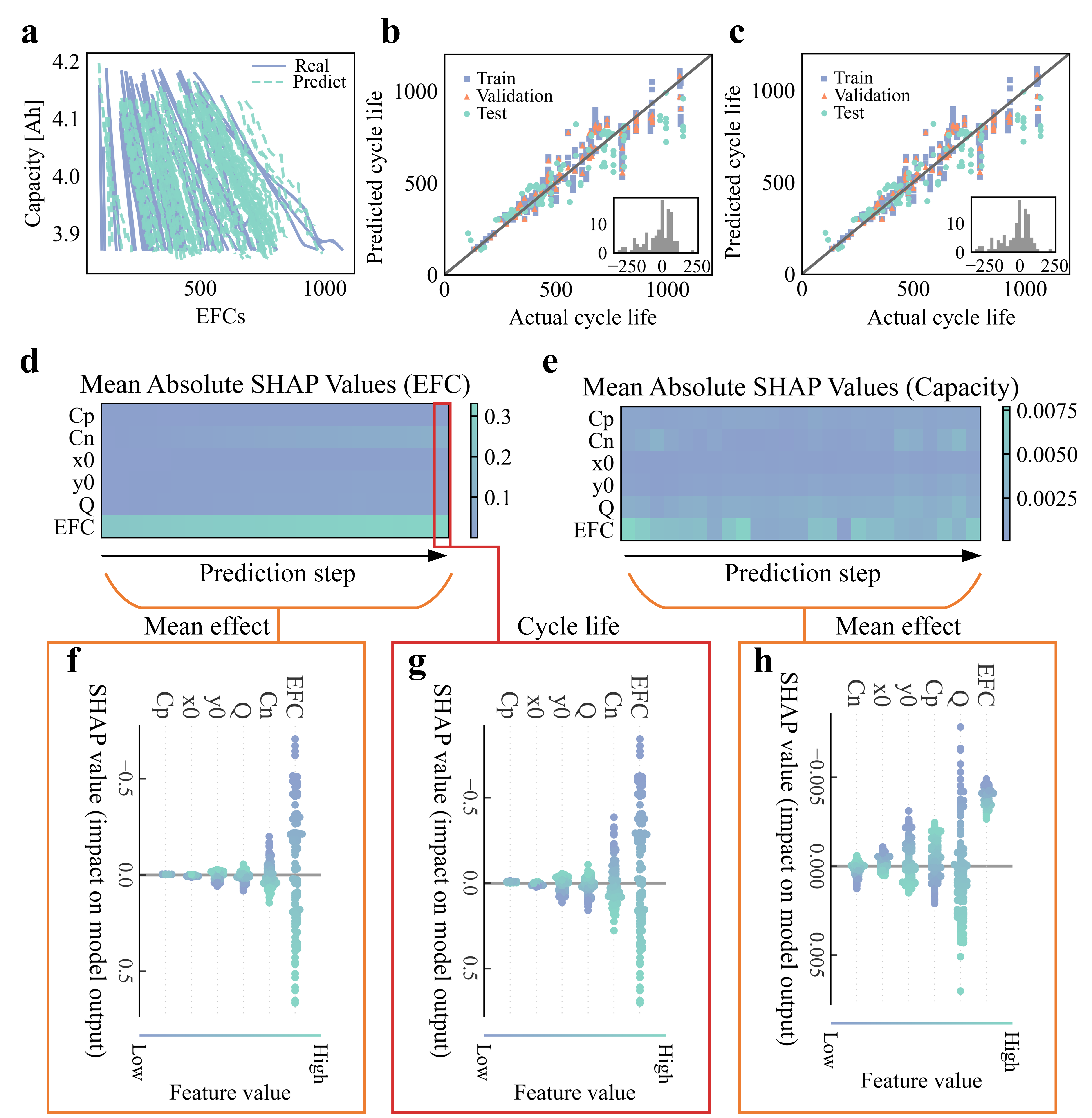}
    \vspace{-0.5cm}
    \caption{\textbf{Future health prognosis.} \textbf{a}) Prediction of future capacity degradation curves. \textbf{b}) Sequence prediction of the cycle life from the last value of the predicted future degradation curve.  \textbf{c}) Point prediction of the cycle life based on the decoder. \textbf{d--e}) The feature impacts analysis of the future degradation predictions based on SHAP for the predicted EFC and capacity values, respectively. \textbf{f}) The mean feature impacts on the predicted future EFC variations. \textbf{g}) The feature impacts on the last prediction step, i.e., the cycle life, of the predicted EFCs. \textbf{h}) The mean feature impacts on the predicted capacities.}
    \label{prog_fig}
    \vspace{0.0cm} 
\end{figure}

The predicted results for the cycle life and future degradation curves are shown in Figs.~\ref{prog_fig} a--c, where the predictions of cycle life (when the capacity C/5 RPT drops below 80\% of the nominal capacity) and future degradation curves are demonstrated. The predictions here show the results when the diagnostic cycle is conducted between 83.5\% and 86.5\% SOH (for an application of guiding onboard predictive maintenance), with earlier prediction results from the third diagnostic cycle presented in Fig.~S17 (for an application of early prediction). The results indicate that the prediction successfully converges to real degradation and predicts the cycle life with mean errors of less than 76 EFCs, 12.8\% of the mean cycle life of 593 EFCs. The predicted cycle life (represented by the EFC) can be determined either from the final value of the future degradation curve (sequence prediction) or directly from the decoder (point prediction), with the comparable performance shown in Table \ref{tab1}. The mean prediction error for early predictions (i.e., predictions from the third diagnostic cycle) is also less than 118 EFCs, and the future capacity degradation curve shows an MAE and root-mean-square error (RMSE) of less than 50.1 and 70.1 mAh, respectively for these cells having nominal capacity of 4.84 Ah. Therefore, the model enables health prognosis for both future degradation curves and cycle life using only onboard measurements, without requiring historical memory. Improved prediction accuracy throughout the aging process supports better predictive maintenance and retirement planning.

To evaluate the feature impacts of the health prognosis decoder, SHAP analysis results are shown in Figs.~\ref{prog_fig}d--h. Each row of the heatmap in Figs.~\ref{prog_fig}de represents the mechanistic and macroscopic features influencing the predicted future degradation curves, described by the variation in capacities and EFCs. The mean impacts on the future degradation curves for the EFCs and capacities are shown in Figs.~\ref{prog_fig}f and \ref{prog_fig}h, respectively, where the current EFC has the highest importance. The anode state has a greater influence on future EFC predictions, while the cathode has a more significant impact on future capacity predictions. The final prediction step of the future EFCs, which represents the cycle life prediction, is analyzed in Fig.~\ref{prog_fig}g, highlighting the substantial impacts of both mechanistic and macroscopic features on the predictions. The SHAP results for prediction of cycle life point, as shown in Fig.~S18, reveal a similar impact of features on the outcomes. When predictions are made at an earlier stage, such as during the third diagnostic cycle as illustrated in Fig.~S17, the feature impacts on future degradation and cycle life predictions exhibit some differences, with mechanistic features playing more prominent roles, especially for the capacity from the anode. This result successfully captures the main aging effects in early life. One physical reason for the differences in the features impacts is that the over-potential of half cells may change during aging, leading to a decreasing effect on mechanistic states extracted in later aging status. In conclusion, the physically constrained health diagnosis model effectively extracts mechanistic states onboard, facilitating a better early health prognosis and cycle life predictions.

\subsection{Applications to Diverse Use Conditions}\label{subsec2-4}

\begin{figure}[H] 
    \centering 
    \includegraphics[width=\textwidth]{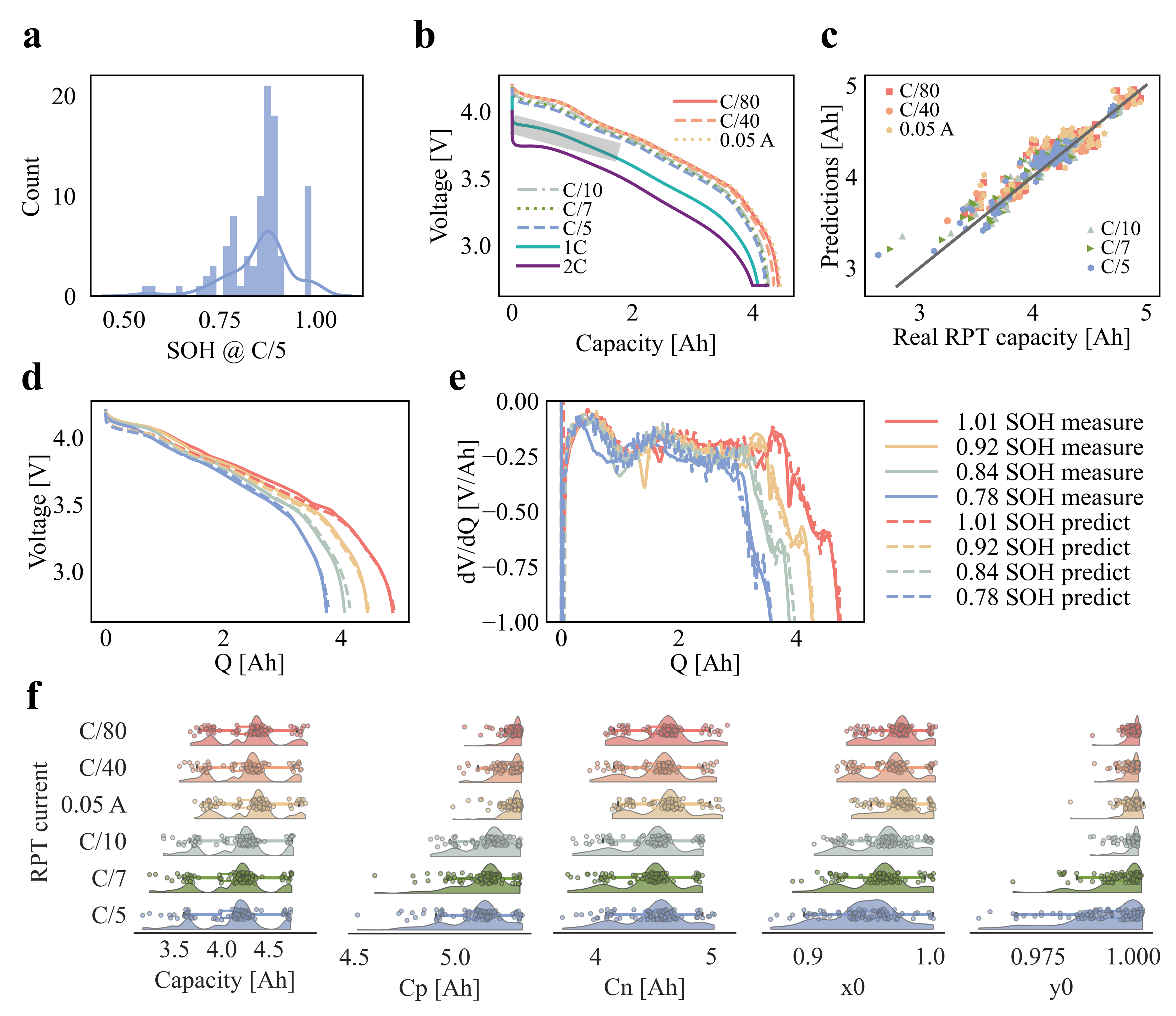}
    \vspace{-0.5cm}
    \caption{\textbf{Rate test validation study.} \textbf{a}) SOH distribution (C/5 discharge) of all the cells. \textbf{b}) Discharge curves with different C-rates. \textbf{c}) Predicted capacity for different C-rate validations using measurement from 1C discharge within 20 min. \textbf{d}) Measured OCV and predicted OCV curves using partial 1 C discharge curves under C/80 at four different SOH levels. \textbf{e}) dV/dQ curves derived from the OCV curves in \textbf{d}. \textbf{f}) Prediction distributions for the discharge capacities using different C-rates and the corresponding constrained mechanistic states.}
    \label{retire_fig}
    \vspace{0.0cm} 
\end{figure}

In the van Vlijmen et al.\ dataset, C/5 discharge data were used as the target pseudo-OCV, however, this is a higher rate than what is typically used in DVA. To address this issue, 94 cells at a variety of SOH (Fig.~\ref{retire_fig}a), calculated by C/5 capacity, were sampled from the van Vlijmen dataset and underwent testing under eight different C-rates with rates as low as C/80 (Fig.~\ref{data_fig}d). Here we deploy the model pre-trained on the van Vlijmen dataset, and fine-tune it to use the partial discharge data from 20 minutes of a 1C discharge as the input to predict the voltage curve and capacity at different C-rates (Fig.~\ref{retire_fig}b). Detailed explanations are provided in Notes S2--S3. The effectiveness of our model for different C-rates validations is illustrated in Fig.~\ref{retire_fig}c, where 5-fold cross-validation was performed, and results for all cells are displayed. Performance of the C/80 pseudo-OCV curve and differential voltage curve predictions for four example cells at different SOH are shown in Figs.~\ref{retire_fig}d-e (SOH based on C/80 capacity). This model demonstrates significant time savings, effectively extracting the low C-rate voltage curves using only 20 minutes of 1C discharge data compared to the nearly 80 hours needed to gather a C/80 pseudo-OCV curve.

Additional results on prediction performance and DVA are provided in Figs.~S19--S20. Fig.~S21 summarizes the prediction outcomes, indicating high accuracy and reliability, with an MAE for predicted capacity and voltage of less than 43.2 mAh and 22.4 mV, respectively, and $R^2$ values exceeding 0.99. Even with a shorter partial discharge, our model maintains strong performance. See Figs.~S22--S23 for results from the 10-minute discharge-based assessment, demonstrating the model's effectiveness with reduced test times. See Fig.~S24 for the feature impact interpretations based on SHAP analysis, where the results also indicate the pronounced impact from anode degradations. By applying a quick and flexible fine-tuning strategy to the model developed for onboard health diagnosis, our model transitions seamlessly to predicting different C-rate behaviors. RPT capacities at different low C-rates can be predicted from partial discharge curves at 1 C, and the mechanistic states that are directly obtained through this process are shown in Fig.~\ref{retire_fig}f. 

\begin{table}[h]
\caption{Summary of the diagnostic and prognostic performances.}\label{tab1}
\begin{tabular*}{\textwidth}{@{\extracolsep\fill}p{1.5cm}ccccc} %
\toprule%
Objective & Task & RMSE & MAE& $R^2$ &  \\
\midrule
\multirow{4}{1.8cm}{Health diagnosis (dataset 1)} 
& Predicted $Q$ of OCV curve ($\hat{Q}$) & 67.2 mAh & 45.1 mAh & 0.998 \\
& Predicted voltage of OCV curve ($\hat{V}$) & 13.8 mV & 10.1 mV & 0.999 \\
& Derived voltage of OCV curve ($V^d$) & 25.2 mV & 18.4 mV & 0.996 \\
& Predicted SOH & 1.86\% & 1.35\% & 0.945 \\
\midrule
\multirow{5}{1.8cm}{Health prognosis (dataset 1)} 
& Predicted $Q$ of future degradation curve & 28.7 mAh & 22.6 mAh & 0.861 \\
& Predicted cycle life of future degradation curve & 62.1 EFC & 39.4 EFC & 0.910 \\
& Predicted cycle life from degradation curve & 105 EFC & 76.0 EFC & 0.824 \\
& Predicted cycle life from model & 105 EFC & 76.4 EFC & 0.824 \\
& Predicted cycle life for all cells & 81.6 EFC & 59.2 EFC & 0.876 \\
\midrule
\multirow{5}{1.8cm}{Early Health prognosis (dataset 1)} %
& Predicted $Q$ of future degradation curve & 70.2 mAh & 50.0 mAh & 0.842 \\
& Predicted EFC of future degradation curve & 93.6 EFC & 60.1 EFC & 0.636 \\
& Predicted cycle life from degradation curve & 158 EFC & 118 EFC & 0.482 \\
& Predicted cycle life from model & 158 EFC & 118 EFC & 0.481 \\
& Predicted cycle life for all cells & 95.9 EFC & 65.1 EFC & 0.794 \\
\midrule
\multirow{4}{1.8cm}{Different C-rates validations (dataset 2)} 
& Predicted $Q$ of OCV curve ($\hat{Q}$) & 68.2 mAh & 43.1 mAh & 0.997 \\
& Predicted voltage of OCV curve ($\hat{V}$) & 27.4 mV & 22.4 mV & 0.996 \\
& Derived voltage of OCV curve ($V^d$) & 52.4 mV & 43.1 mV & 0.984 \\
& Predicted SOH & 2.54\% & 1.86\% & 0.890 \\
\midrule
\multirow{4}{1.8cm}{Dynamic cycling using charging (dataset 3)} 
& Predicted normalized $Q$ of OCV curve ($\hat{Q}$) & 0.61 \% & 0.41 \% & 0.999 \\
& Predicted voltage of OCV curve ($\hat{V}$) & 4.79 mV & 5.05 mV & 0.999 \\
& Derived voltage of OCV curve ($V^d$) & 11.4 mV & 8.46 mV & 0.999 \\
& Predicted SOH & 1.10\% & 0.86\% & 0.963 \\
\midrule
\multirow{4}{1.8cm}{Dynamic cycling using discharging (dataset 3)} 
& Predicted normalized $Q$ of OCV curve ($\hat{Q}$) & 0.79 \% & 0.53 \% & 0.999 \\
& Predicted voltage of OCV curve ($\hat{V}$) & 9.63 mV & 7.33 mV & 0.999 \\
& Derived voltage of OCV curve ($V^d$) & 18.1 mV & 13.7 mV & 0.999 \\
& Predicted SOH & 1.51\% & 1.15\% & 0.932 \\

\botrule
\end{tabular*}
\vspace{-0.1cm}
\footnotetext{Note: dataset 1, 2, and 3 in the table correspond to the van Vlijmen et al. dataset (constant current cycling), the different C-rates dataset (RPT using different C-rates), and the Geslin et al. dataset (dynamic cycling), respectively.}
\end{table}

In real-world applications, although constant current or power may be seen during charging, it is rarely seen during operation or discharge. While the batteries in the Geslin et al.\ dataset are charged identically, they are discharged under several different dynamic discharge loading profiles. We apply our model to the Geslin et al.\  dataset \cite{geslin2024dynamic},validate our model performance under real-world operating conditions. Before using the dynamic discharge data as an input, we first use the charging data as a baseline. The same data sampling technique employed for the van Vlijmen et al.\ dataset is utilized here, where randomly sampled portions of charging curves are extracted from the charging data following the diagnostic cycle for model input, while the C/40 discharge pseudo-OCV serves as the output. The model pre-trained by the Vlijmen dataset is fine-tuned by using approximately one-third of the cells and subsequently tested on the remaining cells. Results are presented in Figs.~S25 and S26 with numerical outcomes detailed in Table~\ref{tab1}. The prediction errors are demonstrated in Fig.~S27 for the predicted OCV and SOH (calculated by C/40 capacity), demonstrating high accuracy and robustness across various dynamic cycling conditions that are closer to practical operation scenarios, with MAE of 5.05 mV and 0.41\% for the predicted voltage and normalized capacity of the OCV curves, and the SOH has an MAE of 0.86\%. Even in this low C-rate OCV case, the predicted OCV has higher accuracy than the derived OCV based on the mechanistic model, again indicating that variations of the half-cell open-circuit curve and other dynamics influence the effectiveness of the mechanistic model. 

\begin{figure}[H] 
    \centering 
    \includegraphics[width=\textwidth]{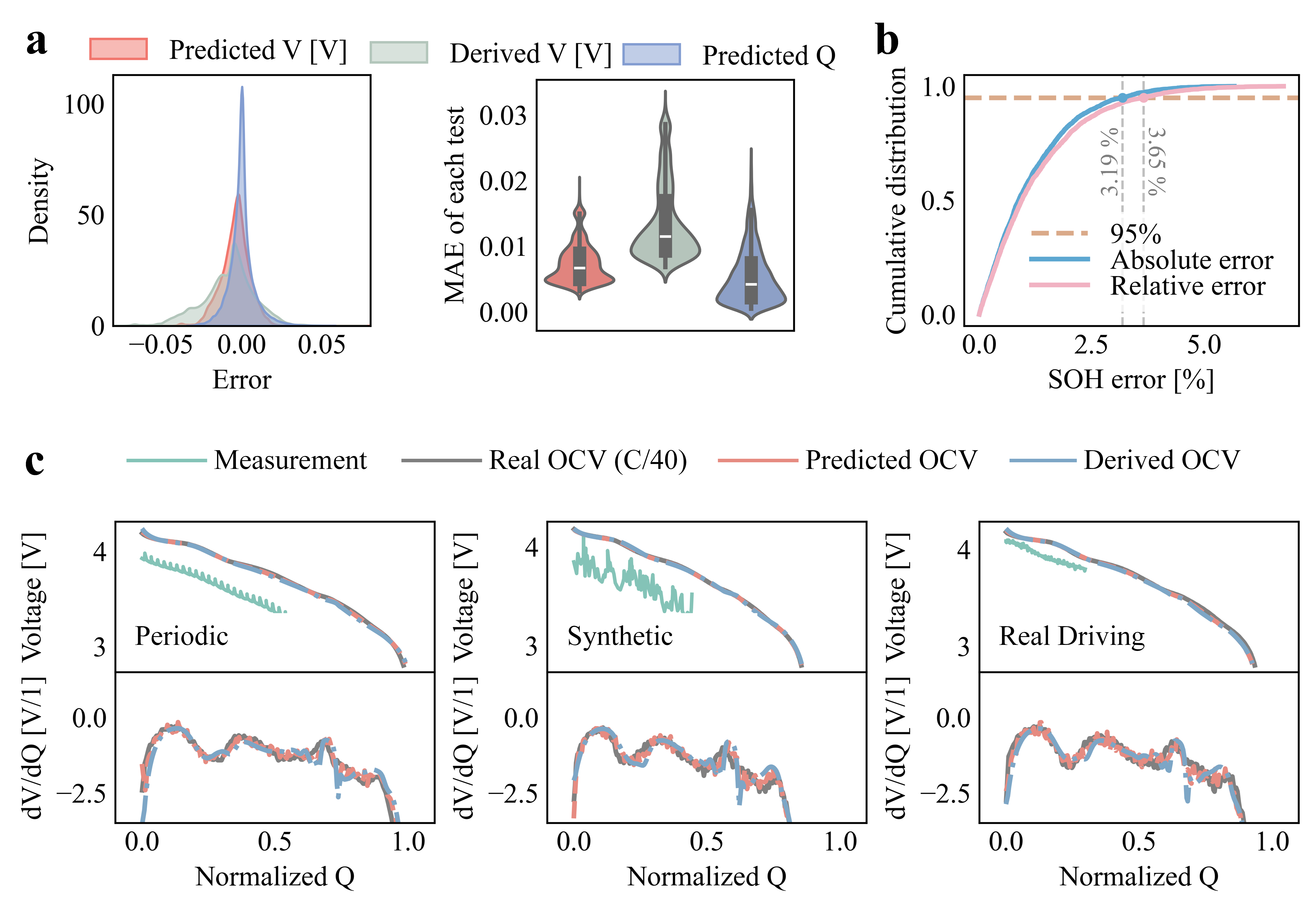}
    \vspace{-0.5cm}
    \caption{\textbf{Validations with dynamic cycling.} The partial discharging data with different loading profiles and C-rates are employed as model input to predict the OCV of the C/40 discharge. \textbf{a}) The error distributions of predicted OCV curves and derived OCV curves for the dynamic cycling dataset. The MAE of each predicted and derived OCV curve of the dynamic cycling dataset. \textbf{b}) The cumulative error distribution of the SOH estimation results. \textbf{c}) Prediction results for three representative cycling conditions including discharging with periodic, synthetic, and real driving profiles respectively.}
    \label{dynamic_fig}
    \vspace{0.0cm} 
\end{figure}

Now, to see if our model can work under practical operating conditions, we instead use randomly sampled portions of the dynamic discharge curves as input for the OCV predictions. The model performance is illustrated in Fig.~\ref{dynamic_fig}, where the overall error distributions for the predicted and derived OCV are presented in Figs.~\ref{dynamic_fig}a. The results indicate good predictions with an MAE of only 7.33 mV and 0.53\% for the voltage and normalized capacity of the predicted OCV curves respectively. The distribution of the accumulated SOH prediction error is shown in Fig.~\ref{dynamic_fig}b, which indicates high prediction accuracy and reliability (95\% of the results have errors less than 3.2\%) under different loading profiles, C-rates, and portion data distributions (more results are shown in Fig.~S28). Three representative prediction results based on truncated dynamic discharge data from periodic, synthetic, and real driving discharging curves are shown in Fig.~\ref{dynamic_fig}c, indicating the effectiveness of our model with different dynamic discharging profiles. Although the results utilizing the identical charging data create more accurate models, we show that reasonable model performance can be achieved even when using the dynamic discharge data. These findings underscore the model's potential for onboard, diagnostic test-free battery health assessment not only under constant current profiles but also dynamic profiles.

\section{Conclusion}\label{sec4}

In summary, we develop an interpretable machine learning model to enable onboard battery health diagnosis and prognosis without time-consuming and disruptive diagnostic tests. We demonstrate the path for combining mechanistic constraints with machine learning leading to a physically interpretable and robust model. To evaluate the versatility of this model, we apply our framework to three battery cycling datasets, encompassing a total of 422 cells. First, we use the van Vlijmen et al.\ dataset to reconstruct C/5 diagnostic cycle data from constant current cycling across various state-of-charge windows. Next, we generate a new rate-performance dataset by testing rates as low as C/80 with 94 degraded cells from the van Vlijmen et al.\ study. Using this dataset, we fine-tune the model to reconstruct information from low-rate performance tests (such as C/80 and C/40), leveraging partial data from a 1C discharge. Finally, we fine-tune the model to reconstruct C/40 diagnostic cycle data using partial dynamic operating condition data from the Geslin et al.\ dataset. Once the model is trained or fine-tuned to a given system, diagnostic-free onboard battery health assessment is possible. By applying this framework across these three datasets, we demonstrate its utility in enabling diagnostic-free onboard battery diagnosis and prognosis.

Our model can be flexibly applied across broad application interests such as materials and mechanisms evaluation, synthetic data generation, electric transportation health assessment, and fast retiring assessment, with slight fine-tuning. We encourage future work in this field to further understand the detrimental effect of diagnostic cycles on battery aging, and to utilize diagnostic-free battery health monitors to circumvent these issues.

\section{Methods}\label{sec5}
\subsection{Experiment and data generation}\label{subsec5-1}
236 batteries with cathode of nickel cobalt aluminum (NCA) and anode of graphite-SiO$_{x}$ from \cite{van2023interpretable} are selected in the first dataset. The cells aged below 80\% and with available raw data are used in this work. More details of the experiment description are found in \cite{van2023interpretable}, and all the data is processed through the BEEP processing pipeline \cite{herring2020beep}. 94 batteries are collected for the different C-rates validation tests using the same test platform to generate the second dataset. The batteries are 21700 cylindrical cells produced by Panasonic with a nominal capacity of 4.84 Ah. These 94 cells are aged until different SOH levels through different cycling profiles. Eight different C-rates ranging from C/80 to 2C were used for the RPT tests. Finally, the dynamic cycling dataset from Geslin et al.\  \cite{geslin2024dynamic} is utilized to validate the proposed model, which is designed for real-world aging conditions. 92 commercial silicon oxide--graphite/nickel cobalt aluminum cells were cycled. For more comprehensive details, readers are referred to \cite{geslin2024dynamic}.

\subsection{Machine learning with mechanistic constraints}\label{subsec5-2}

Auto-encoders and decoders with physical constraints are designed for the machine learning model, where the mechanistic model describing the relationship between the four key states and the full-cell OCV is constrained to make the data-driven model interpretable. The auto-encoders and decoders can be designed as different types of neural networks while multilayer perceptron is simpler and easier to explain compared to others. In this paper, the mechanistic constraints are based on the DVA. A similar method presented in previous works \cite{marongiu2016board, dubarry2012synthesize, stadlercombining, lin2024identifiability, Lee_2020} is employed here. The mechanistic model is to fit the full-cell OCV by means of two half-cell open circuit potentials, named OCP$_\text{p}$ for the cathode and OCP$_\text{n}$ for the anode. During battery aging, four key mechanistic states $\theta = \left[ x_0, y_0, C_\text{p}, C_\text{n} \right]$ need to be extracted to represent the shift and shrinking of the OCP curves and describe the aging mechanisms. In this way, the non-destructive aging mode diagnosis can be performed and the LLI and LAM at two electrodes can be obtained. Detailed descriptions of the model are given in Note S4. Instead of using offline fitting for the OCV curve to prepare labels of the four mechanistic states, we construct a constraint between the output of the decoder and the states encoded by the encoder through the mechanistic model equations described in Note S4. In addition, the prior knowledge of the boundary condition for the prediction and constrained states helps maintain the predictions within reasonable ranges and accelerates the model convergence. Therefore, the total loss function for the health diagnosis can be described by
\begin{equation}
L_1 = \alpha L_{\text{reg}} + \beta L_{\text{phy}} + \sum_{i=1}^{N} \gamma_i L_{\text{bound},i},
\end{equation}
where $L_{\text{reg}}$, $L_{\text{phy}}$, and $L_{\text{bound}}$ represent the loss for the prediction, physical constraints, and boundary constraints, respectively; and $\alpha$, $\beta$, $\gamma$ are the associated weighting factors. For prognosis, the regression loss and boundary knowledge are included for the prognosis decoder training, and the total loss is
\begin{equation}
L_2 = \eta L_{\text{reg}} + \sum_{i=1}^{N} \zeta_i L_{\text{bound},i},
\end{equation}
where $\eta$ and $\zeta$ also are the weighting factors. The model is developed through PyTorch. To understand the feature impacts on the diagnosis and prognosis performance, we use SHAP python library of the encoder and decoder respectively. See Notes S3 and S6 for detailed descriptions of the machine learning model and SHAP analysis \cite{NIPS2017_8a20a862, mangalathu2020failure}. 

\backmatter

\section{Data Availability}
Data used in this work will be available after publication.

\section{Acknowledgements}
This work was supported by the Toyota Research Institute through the Accelerated Materials Design and Discovery program. Y.C. acknowledges the Independent Research Foundation Denmark and Novo Nordisk Foundation. The authors would like to thank A. Geslin and X. Cui for their thoughtful comments and feedback on the paper.

\bibliography{sn-bibliography}

\end{document}